\input harvmac

%\draftmode

\def\inbar{\,\vrule height1.5ex width.4pt depth0pt}
\def\IZ{\relax\ifmmode\mathchoice
{\hbox{\cmss Z\kern-.4em Z}}{\hbox{\cmss Z\kern-.4em Z}}
{\lower.9pt\hbox{\cmsss Z\kern-.4em Z}} {\lower1.2pt\hbox{\cmsss
Z\kern-.4em Z}}\else{\cmss Z\kern-.4em Z}\fi}
\def\IB{\relax{\rm I\kern-.18em B}}
\def\IC{{\relax\hbox{$\inbar\kern-.3em{\rm C}$}}}
\def\ID{\relax{\rm I\kern-.18em D}}
\def\IE{\relax{\rm I\kern-.18em E}}
\def\IF{\relax{\rm I\kern-.18em F}}
\def\IG{\relax\hbox{$\inbar\kern-.3em{\rm G}$}}
\def\IGa{\relax\hbox{${\rm I}\kern-.18em\Gamma$}}
\def\IH{\relax{\rm I\kern-.18em H}}
\def\II{\relax{\rm I\kern-.18em I}}
\def\IK{\relax{\rm I\kern-.18em K}}
\def\IP{\relax{\rm I\kern-.18em P}}
\def\IQ{\relax\hbox{$\inbar\kern-.3em{\rm Q}$}}

\def\PP{\IP}
\def\CC{\IC}
\font\cmss=cmss10 \font\cmsss=cmss10 at 7pt
\def\rlx{\relax\leavevmode}
\def\ZZ{\rlx\leavevmode\ifmmode\mathchoice{\hbox{\cmss Z\kern-.4em
Z}}{\hbox{\cmss Z\kern-.4em Z}}{\lower.9pt\hbox{\cmsss Z\kern-.36em
Z}} {\lower1.2pt\hbox{\cmsss Z\kern-.36em Z}}\else{\cmss Z\kern-.4em
Z}\fi}

\def\frac#1#2{{#1 \over #2}}

\def\a{\alpha}

\def\p{\phi}
\def\r{\rho}
\def\th{\theta}
\def\bth{{\bar \theta}}

\def\ba{{\bar a}}
\def\bb{{\bar b}}
\def\bc{{\bar c}}

\def\bz{{\bar z}}
\def\pa{{\partial}}
\def\bpa{\bar{\partial}}
\def\zo{z_0}
\def\za{z_1}
\def\zb{z_2}
\def\zc{z_3}
\def\zd{z_4}
\def\bza{\bar z_1}
\def\bzb{\bar z_2}

\def\Za{Z_1}
\def\Zb{Z_2}
\def\Zc{Z_3}
\def\Zd{Z_4}
\def\xo{x_0}
\def\xa{x_1}
\def\yo{y_0}
\def\ya{y_1}
\def\Xo{{X'}}
\def\Yo{{Y'}}
\def\om{\omega}
\def\Om{\Omega}
\def\w{\wedge}

\def\rk{{\rm rk}\,}
\def\Nk{{N(\rho)}}
\def\Nkc{N(\rho)^\perp}
\def\Tr{{\rm Tr}}
\def\z{\zeta}

%%%%%%%%%%%%%%%%%%%%%%%%%%%%%%%%%%%%%%%%%%%%%%%%%%%%%%%%%%%%%%%%%%%%%%

%\cite{Braun:2005ux}
\lref\smI{V.~Braun, Y.~H.~He, B.~A.~Ovrut and T.~Pantev, ``A
heterotic standard model,''
  Phys.\ Lett.\  B {\bf 618}, 252 (2005)
  [arXiv:hep-th/0501070].}
  %%CITATION = PHLTA,B618,252;%%

%\cite{Bouchard:2005ag}
\lref\smII{V.~Bouchard and R.~Donagi, ``An SU(5) heterotic standard
model,'' Phys.\ Lett.\  B {\bf 633}, 783 (2006)
[arXiv:hep-th/0512149].}
  %%CITATION = PHLTA,B633,783;%%

%\cite{Cleaver:1998saa}
\lref\cfn{G.~B.~Cleaver, A.~E.~Faraggi and D.~V.~Nanopoulos,
``String derived MSSM and M-theory unification,''
  Phys.\ Lett.\  B {\bf 455}, 135 (1999)
  [arXiv:hep-ph/9811427].}
  %%CITATION = PHLTA,B455,135;%%

%\cite{Beasley:2008dc}
\lref\bhv{
  C.~Beasley, J.~J.~Heckman and C.~Vafa,
  ``GUTs and exceptional branes in F-theory - I,''
  [arXiv:0802.3391].}
  %%CITATION = ARXIV:0802.3391;%%

%
%\cite{Donagi:2008ca}
\lref\dowi{
  R.~Donagi and M.~Wijnholt,
  ``Model building with F-theory,''
  [arXiv:0802.2969].}
  %%CITATION = ARXIV:0802.2969;%%

\lref\MoVa{D.~Morrison and C.~Vafa, ``Compactifications of F-theory
on Calabi-Yau threefolds- II," [arXiv:hep-th/9603161].}

\lref\yau{S.-T. Yau, ``Compact three dimensional K\"ahler manifolds
with zero Ricci curvature," in {\it Symposium on anomalies,
geometry, and topology}, ed. William A. Bardeen and Alan R. White
(1985) 395.}

\lref\GSW{M.~B.~Green, J.~H.~Schwarz and E.~Witten, {\it
Superstring theory: volume 2}, Cambridge University Press (1987).}

%\cite{Candelas:1985en}
\lref\vc{P.~Candelas, G.~T.~Horowitz, A.~Strominger and E.~Witten,
``Vacuum configurations for superstrings,'' Nucl.\ Phys.\  B {\bf
258}, 46 (1985).}
  %%CITATION = NUPHA,B258,46;%%

\lref\hull{C.~M.~Hull, ``Superstring compactifications with torsion and
space-time supersymmetry," in {\it 1st Torino meeting on superunification and
extra dimensions, September 1985, Torino, Italy}, World Scientific, R.~D'Auria and D.~Fre,
 editors; Singapore, 347 (1986);
  ``Anomalies, ambiguities and superstrings,''
  Phys.\ Lett.\  B {\bf 167}, 51 (1986);
  %%CITATION = PHLTA,B167,51;%%
  ``Compactifications of the heterotic superstring,''
  Phys.\ Lett.\  B {\bf 178}, 357 (1986).
  %%CITATION = PHLTA,B178,357;%%
}

\lref\FY{J.-X.~Fu and S.-T.~Yau, ``The theory of superstring with
flux on non-K\"ahler manifolds and the complex Monge-Amp\'ere
equation,'' J.\ Differential\ Geom.\  {\bf 78}, 369 (2008) [arXiv:hep-th/0604063].
%%CITATION = HEP-TH 0604063;%%
}

\lref\BBFTY{
  K.~Becker, M.~Becker, J.-X.~Fu, L.-S.~Tseng and S.-T.~Yau,
   ``Anomaly cancellation and smooth non-K\"ahler solutions in heterotic string theory,''
  Nucl.\ Phys.\ B {\bf 751}, 108 (2006)
  [arXiv:hep-th/0604137].}
  %%CITATION = HEP-TH 0604137;%%

\lref\adel{
  A.~Adams, M.~Ernebjerg and J.~M.~Lapan,
  ``Linear models for flux vacua,''
  Adv.\ Theor.\ Math.\ Phys.\ {\bf 12}, 821 (2008)
  [arXiv:hep-th/0611084].}
  %%CITATION = HEP-TH 0611084;%%}

\lref\kiyi{
  T.~Kimura and P.~Yi,
  ``Comments on heterotic flux compactifications,''
  JHEP {\bf 0607}, 030 (2006)
  [arXiv:hep-th/0605247].}
  %%CITATION = HEP-TH 0605247;%%

\lref\cylap{
  M.~Cyrier and J.~M.~Lapan,
``Towards the massless spectrum of non-Kaehler heterotic
compactifications,'' Adv.\ Theor.\ Math.\ Phys.\ {\bf 10}, 853 (2007)
  [arXiv:hep-th/0605131].
  %%CITATION = 00203,10,853;%%
}

\lref\bty{M.~Becker, L.-S.~Tseng and S.-T.~Yau,
  ``Heterotic Kahler/non-Kahler transitions,'' Adv.\ Theor.\ Math.\ Phys.\  {\bf 12}, 1151 (2008)
  [arXiv:0706.4290].}
  %%CITATION = ARXIV:0706.4290;%%

%\cite{Strominger:1986uh}
\lref\strom{A.~Strominger,  ``Superstrings with torsion,''
Nucl.\ Phys.\ B {\bf 274}, 253 (1986).}
  %%CITATION = NUPHA,B274,253;%%

%\cite{deWit:1986xg}
\lref\dewit {B.~de Wit, D.~J.~Smit and N.~D.~Hari Dass,
``Residual supersymmetry of compactified d=10 supergravity'' Nucl.\
Phys.\ B {\bf 283}, 165 (1987).}
  %%CITATION = NUPHA,B283,165;%%

%\cite{Becker:2003yv}
\lref\beck{K.~Becker, M.~Becker, K.~Dasgupta and P.~S.~Green,
``Compactifications of heterotic theory on non-Kaehler complex
manifolds I,'' JHEP {\bf 0304}, 007 (2003), [arXiv:hep-th/0301161].}
  %%CITATION = HEP-TH 0301161;%%

%\cite{Becker:2003sh}
\lref\becktwo{K.~Becker, M.~Becker, P.~S.~Green, K.~Dasgupta and
E.~Sharpe, ``Compactifications of heterotic strings on
non-Kaehler complex  manifolds II,'' Nucl.\ Phys.\ B {\bf 678}, 19
(2004), [arXiv:hep-th/0310058].}
  %%CITATION = HEP-TH 0310058;%%

%\cite{Adams:2001sv}
\lref\tachyonI{
  A.~Adams, J.~Polchinski and E.~Silverstein,
``Don't panic! Closed string tachyons in ALE space-times,''
  JHEP {\bf 0110}, 029 (2001)
  [arXiv:hep-th/0108075].}
  %%CITATION = JHEPA,0110,029;%%

%\cite{Horowitz:2007pr}
\lref\tachyonII{
  G.~T.~Horowitz, J.~Orgera and J.~Polchinski,``Nonperturbative
  instability of $AdS_5 \times S^5/Z_k$,''
  Phys.\ Rev.\  D {\bf 77}, 024004 (2008)
  [arXiv:0709.4262 [hep-th]].}
  %%CITATION = PHRVA,D77,024004;%%

%\cite{Kachru:1998ys}
\lref\kasi{S.~Kachru and E.~Silverstein, ``4d conformal theories and
strings on orbifolds,''
  Phys.\ Rev.\ Lett.\  {\bf 80}, 4855 (1998)
  [arXiv:hep-th/9802183].}
  %%CITATION = PRLTA,80,4855;%%

%\cite{Dasgupta:1999ss}
\lref\rds{K.~Dasgupta, G.~Rajesh and S.~Sethi, ``M theory,
orientifolds and G-flux,'' JHEP {\bf 9908}, 023 (1999)
  [arXiv:hep-th/9908088].}
  %%CITATION = JHEPA,9908,023;%%

%\cite{Becker:2002sx}
\lref\bd{K.~Becker and K.~Dasgupta, ``Heterotic strings with
torsion,'' JHEP {\bf 0211}, 006 (2002)
  [arXiv:hep-th/0209077].}
  %%CITATION = JHEPA,0211,006;%%

\lref\hls{
  J.~A.~Harvey, D.~A.~Lowe and A.~Strominger,
  ``N=1 string duality,''
  Phys.\ Lett.\  B {\bf 362}, 65 (1995)
  [arXiv:hep-th/9507168].
  %%CITATION = PHLTA,B362,65;%%
}

%\cite{Vafa:1995gm}
\lref\vawi{C.~Vafa and E.~Witten, ``Dual string pairs with N = 1 and
N = 2 supersymmetry in four  dimensions,''
  Nucl.\ Phys.\ Proc.\ Suppl.\  {\bf 46}, 225 (1996)
  [arXiv:hep-th/9507050].}
  %%CITATION = NUPHZ,46,225;%%

\lref\hit{
 N.~Hitchin, ``Compact four-dimensional Einstein
manifolds," J.\ Diff.\ Geometry, {\bf 9} 435 (1974). }

\lref\MOP{ D.~G.~Markushevich, M.~A.~Olshanetsky, and
A.~M.~Perelomov, ``Description of a class of superstring
compactifications related to semi-simple Lie algebras," Commun.\
Math.\ Phys.\ {\bf 111} 247 (1987).}

\lref\RoanYau{S.-S.~Roan and S.-T.~Yau, ``On Ricci flat 3-fold,"
Acta\ Math.\ Sinica\ {\bf 3} No. 3 256 (1987)}

\lref\StrWi{A.~Strominger and E.~Witten, ``New manifolds for
superstring compactification,'' Commun.\ Math.\ Phys.\  {\bf 101},
341 (1985).}
  %%CITATION = CMPHA,101,341;%%

\lref\Aspinwall{P.~S.~Aspinwall, ``K3 surfaces and string duality,''
[hep-th/9611137].}
  %%CITATION = HEP-TH/9611137;%%

\lref\Hubsch{T.~H\"ubsch, {\it Calabi-Yau manifolds: a bestiary for
physicists}, World Scientific, New Jersey (1992).}

\lref\BSV{M.~Bershadsky, V.~Sadov and C.~Vafa, ``D-strings on
D-manifolds," [arXiv:hep-th/9510225].}

\lref\orbi{L.~J.~Dixon, J.~A.~Harvey, C.~Vafa and E.~Witten,
  ``Strings on orbifolds,''
  Nucl.\ Phys.\  B {\bf 261}, 678 (1985); L.~J.~Dixon, J.~A.~Harvey, C.~Vafa and E.~Witten,
  ``Strings on orbifolds. 2,''
  Nucl.\ Phys.\  B {\bf 274}, 285 (1986).
  %%CITATION = NUPHA,B274,285;%
  }

\lref\gp{E.~Goldstein and S.~Prokushkin,
  ``Geometric model for complex non-K\"ahler manifolds with SU(3) structure,''
  Commun.\ Math.\ Phys.\  {\bf 251}, 65 (2004)
  [arXiv:hep-th/0212307].
  %%CITATION = CMPHA,251,65;%%
}

\lref\arts{M.~Artebani and A.~Sarti, ``Non-symplectic  automorphisms
of order $3$ on $K3$ surfaces," [arXiv:0801.3101].}

\lref\taki{S.~Taki, ``Classification of non-symplectic automorphisms
or order $3$ on $K3$ surfaces," [arXiv:0802:1956].}

%\lref\zhang{D.-Q.~Zhang, ``Logarithmic Enriques surfaces," J.\
%Math.\ Kyoto\ Univ., {\bf 31} 419 (1991).}

\lref\nika{V.~V.~Nikulin, ``Factor groups of groups of automorphisms
of hyperbolic forms with respect to subgroups generated by
$2$-reflections," J.\ Soviet\ Math., {\bf 22} 1401 (1983).}

\lref\alexnik{V.~Alexeev and V.~V.~Nikulin, {\it Del Pezzo and $K3$
surfaces}, Mathematical Society of Japan, Tokyo (2006).}

\lref\Greene{
  B.~R.~Greene,
  ``String theory on Calabi-Yau manifolds,''
  [arXiv:hep-th/9702155].
  %%CITATION = HEP-TH/9702155;%%
}

\lref\bese{Talk of K.~Becker at {\it String Phenomenlogy 2008}, May
28 - June 1, 2008, University of Pennsylvania; K~Becker and
S.~Sethi, paper to appear.}

\lref\unpub{M.~Becker, L.-S.~Tseng and S.~T.~Yau, unpublished.}

\lref\CE{E.~Calabi and B.~Eckmann, ``A class of compact,
complex manifolds which are not algebraic" Ann.\ of Math.\ {\bf 58}, 494 (1953).}

\lref\KM{
K.~Becker and M.~Becker,
``M-theory on eight-manifolds,''
 Nucl.\ Phys.\  B {\bf 477}, 155 (1996)
 [arXiv:hep-th/9605053].
  %%CITATION = NUPHA,B477,155;%%
}

\lref\FTY{J.-X.~Fu, L.-S.~Tseng and S.-T.~Yau,
  ``Local heterotic torsional models,''
  [arXiv:0806. 2392].
  %%CITATION = ARXIV:0806.2392;%%
}

\lref\Morris{D.~R.~Morrison, ``On K3 surfaces with large Picard number," Invent.\ Math.\ {\bf 75}, (1984) 105.}

\lref\Calabi{E.~Calabi, ``M\'etriques K\"ahl\'eriennes et fibr\'es holomorphes," Ann.\ Sci.\ Ec.\ Norm.\ Super.\ {\bf 12}, (1979) 269.}

\lref\andreas{B.~Andreas and G.~Curio,
  ``Heterotic models without fivebranes,''
  J.\ Geom.\ Phys.\  {\bf 57}, 2136 (2007)
  [arXiv:hep-th/0611309];
  %%CITATION = JGPHE,57,2136;%%
 ``Deformations of bundles and the standard model,''
  Phys.\ Lett.\  B {\bf 655}, 290 (2007)
  [arXiv:0706.1158];
  %%CITATION = PHLTA,B655,290;%%
 ``Extension bundles and the standard model,''
  JHEP {\bf 0707}, 053 (2007)
  [arXiv:hep-th/0703210].
  %%CITATION = JHEPA,0707,053;%%
}

\lref\Card{
  G.~L.~Cardoso, G.~Curio, G.~Dall'Agata, D.~Lust, P.~Manousselis and G.~Zoupanos,
  ``Non-K\"ahler string backgrounds and their five torsion classes,''
  Nucl.\ Phys.\  B {\bf 652}, 5 (2003)
  [arXiv:hep-th/0211118].
  %%CITATION = NUPHA,B652,5;%%
}

\lref\GMW{
  J.~P.~Gauntlett, D.~Martelli and D.~Waldram,
  ``Superstrings with intrinsic torsion,''
  Phys.\ Rev.\  D {\bf 69}, 086002 (2004)
  [arXiv:hep-th/0302158].
  %%CITATION = PHRVA,D69,086002;%%
}

%%%%%%%%%%%%%%%%%%%%%%%%%%%%%%%%%%%%%%%%%%%%%%%%%

\Title{ {\vbox{
%\rightline{\hbox{hep-th/yymmnnn}
\rightline{\hbox{MIFP-08-03}}
}}} {\vbox{ \hbox{\centerline{ New
Heterotic Non-K\"ahler Geometries}}\hbox{} \hbox{\centerline{}} }}
%centerline{\bf Preliminary Draft}

\bigskip\bigskip

\centerline{Melanie Becker$^{1}$, ~Li-Sheng
Tseng$^{2,3}$~and~Shing-Tung Yau$^2$}

\bigskip \bigskip

\centerline{$^1$ \it George P. and Cynthia W. Mitchell Institute for
Fundamental Physics}

\centerline{\it Texas A \& M University, College Station, TX 77843,
USA}

\medskip

\centerline{$^2$\it Department of Mathematics, Harvard University,
Cambridge, MA 02138, USA}

\medskip

\centerline{$^3$\it Center for the Fundamental Laws of Nature}

\centerline{\it Jefferson Physical Laboratory, Harvard University,
Cambridge, MA 02138, USA}

\bigskip
\bigskip

\bigskip

\centerline{\bf Abstract}

\bigskip
New heterotic torsional geometries are constructed as orbifolds of
$T^2$ bundles over $K3$. The discrete symmetries considered can be
freely-acting or have fixed points and/or fixed curves. We give
explicit constructions when the base $K3$ is Kummer or algebraic.
The orbifold geometries can preserve ${\cal N}=1,2$ supersymmetry in
four dimensions or be non-supersymmetric.

%The resolution of the singular models may result in low
%energy models with a small number of net generations.
%as well as stablized moduli fields.
%An algebraic description of torsional geometries can be given for base
%$K3$'s which are algebraic. %branched covering description of $K3$ and projective $K3$ surfaces
%we develop an algebraic description of these geometries.

\bigskip
\baselineskip 18pt
\bigskip
\noindent

\Date{June, 2008}

%%%%%%%%%%%%%%%%%%%%%%%%%%%%%%%%%%%%%%%%%%%%%%%

\newsec{Introduction}

Heterotic string compactifications play a prominent role in string
theory model building as the standard model gauge group can be
easily incorporated in this approach. Some recent progress in
heterotic model building was made in \smII, where a
heterotic M-theory model was constructed that reproduces the charged
particle content of the MSSM without additional exotics.\foot{Other standard model motivated heterotic constructions are given in \refs{\smI, \andreas}.   A CFT
model with no exotics has also been presented in \cfn.}  Interestingly, F-theory
models that incorporate GUT gauge groups have also recently been
constructed in the literature \refs{\dowi, \bhv}  Common to these
recent developments is that moduli fields coming either from metric
deformations and/or deformations describing brane positions are not
stabilized, so that making predictions for physical observables
becomes challenging.

Now more than ever it is important to develop techniques to
construct heterotic flux compactifications in which moduli
stabilization can be addressed.  Having a heterotic three generation
model with stabilized moduli allows us to remove this (uncharged)
``exotic'' matter and in principle predict the masses of quarks and
leptons, an important step towards connecting string theory to the
real world.

Heterotic flux compactifications have been known for quite some
time, starting with the seminal works of \refs{\strom, \hull,
\dewit} in the mid-1980's.  All the more, it is surprising that the
number of torsional geometries that can serve as backgrounds for
heterotic string compactifications seems rather limited.  The most
studied class of
 smooth heterotic torsional compactifications is the FSY geometry \refs{\FY, \BBFTY}.
  The manifold is a $T^2$ bundle over a $K3$ surface \refs{\gp,\beck} similar to the
  well-known non-K\"ahler torus bundle construction of Calabi and Eckman \CE.
   The solution was first identified as the heterotic dual of M-theory on $K3\times K3$
   with non-zero $G$-flux \refs{\rds,\bd}.  It was proven in \FY\ to satisfy both
   the requirements of supersymmetry and the anomaly condition of the heterotic theory.
   A conformal field theory description in the language of a gauged linear sigma model
   has also been developed in \adel.

As was discussed in \refs{\beck, \becktwo}, this model has a vanishing Euler
 characteristic  as well as a vanishing number of generations. When
the spin connection is embedded into the gauge connection, the net
number of generations is given by the Euler characteristic of the
internal Calabi-Yau geometry \vc. However, for the more general
heterotic flux compactifications that we are interested in, where the
spin connection cannot be embedded into the gauge connection, the
net number of generations is given by one-half of the third Chern number of the
gauge bundle \becktwo.

Our goal here is to expand the set of consistent heterotic torsional geometries by
constructing orbifolds of the FSY geometry.  Orbifolding techniques \orbi\ were
already used in the early days of string model building to construct
models with a small number of generations. Modding out some
heterotic string compactifications with an unrealistically large
number of generations by a discrete symmetry, led to more realistic
models (see {\it e.g.} \refs{\yau,\StrWi}).  In fact, the first heterotic
three generation model was constructed by quotienting a complete
intersection Calabi-Yau three-fold \yau.

To construct orbifolds of a $T^2$ bundle over $K3$ surface, we use special
classes of $K3$ base surfaces. One set of models arises from Kummer $K3$ surfaces
which have a $T^4/\IZ_2$ orbifold limit.  A second set of models uses algebraic $K3$
surfaces with finite discrete group actions.  The simplest ones are those with
a branched covering description. The orbifold actions described in this approach
are either freely acting or have fixed curves and/or fixed points.  The freely-acting
 orbifold actions give smooth geometries which may preserve ${\cal N}=0,1,2$ supersymmetry
  in four dimensions.  For the non-freely acting discrete symmetries, the resulting
  orbifolds will contain singularities which geometrically need to be resolved.
    Physically, if the resolution can involve turning on non-vanishing gauge bundles,
     this may lead to models with a non-zero third Chern class.  Unfortunately,
     unlike the Calabi-Yau case where orbifold singularities can be straightforwardly
     resolved by means of toric resolution (see reviews
     in \refs{\Hubsch, \Aspinwall, \Greene}), singularity resolution in
      the non-K\"ahler context seems much more involved and has not been studied
      previously.  The required analysis is delicate and rather technical, and hence,
      we will leave the discussion of non-K\"ahler resolution for a subsequent work.

The organization of this paper is as follows. In Section 2, we review
the properties of torsional manifolds that we need in the later
sections and describe the FSY geometry in some detail.  In Section
3, we discuss properties of the orbifolds of the FSY geometry, {\it i.e.}
 $Y=X/\Gamma$ where $\Gamma$ is a discrete symmetry group of $X$, a torus bundle over
  $K3$ geometry.  The discrete symmetries can act on the torus fiber either as a shift
  or a rotation. In the former case the orbifold actions are always freely acting and
   lead to smooth models with vanishing Euler characteristic and third Chern class.
   In the latter, the orbifold action may have fixed points and fixed curves singularities.
   In Section 4, we construct models that realize these different possibilities starting
   from the FSY geometry with the $K3$ base in the $\IZ_2$ orbifold limit $T^4/\IZ_2$.
In Section 5,  we take as the base $K3$ those with a branched covering description to
build a second set of orbifold torsional geometries.  Section 6
discusses constructions based on more general algebraic $K3$ base with a discrete symmetry group action.  In particular, we give a supersymmetric orbifold model with only fixed point singularities.   Orbifold models with only fixed points should be easier to resolve than those
  containing fixed curves.  In Section 7, we present our conclusions and raise some open questions.

\newsec{Review of Torsional Geometry}

The internal six-dimensional complex geometries of supersymmetric
flux compactifications of the heterotic string are characterized by
a hermitian $(1,1)$-form $J$, a holomorphic $(3,0)$-form $\Omega$,
and a non-abelian gauge field strength $F$ that are constrained by
supersymmetry and anomaly cancellation conditions \strom
\eqn\balance{d(\|\Omega\|_J\, J\w J)=0~,}
\eqn\hermitym{F^{(2,0)}=F^{(0,2)}=0~,\quad F_{mn}J^{mn}=0~,}
\eqn\anomcan{2i\, \pa\bpa J = {\a' \over 4} [{\rm tr} (R\w R) - {\rm
tr}(F\w F)]~.} These equations (sometimes termed the ``Strominger
system'' in the mathematics literature) provide the necessary and
sufficient conditions for space-time supersymmetry in four
dimensions and imply the equations of motion to one-loop order in
$\alpha'$.  Given a solution, the physical fields $(g, H_3, \phi)$
are expressed in terms of $(J,\Omega)$ as \eqn\jrel{g_{a\bb}=-i
J_{a\bb}~,} \eqn\Hrel{H_{ab\bc}=-i(\pa_a J_{b\bc} - \pa_b
J_{a\bc})~,} \eqn\prel{e^{-4\p} = \frac{i}{3!}\Om_{abc}{\bar
\Om}_{\ba\bb\bc}J^{a\ba}J^{b\bb}J^{c\bc}~,} where we have set the
integrable complex structure to take the diagonal form
$J_a{}^b=i\,\delta_a{}^b$.

The most well-known solution to these equations consists of the
complex $T^2$ bundle over a $K3$ surface which we denote by $X$.\foot{A $T^4$ base geometry can satisfy the necessary $SU(3)$ intrinsic torsion conditions of conformally balanced \balance\ and the existence of a holomorphic three-form \Card.  It would however not be consistent with the anomaly condition \refs{\GMW,\BBFTY}.}  The explicit form of the holomorphic $(3,0)$-form for this manifold is
\eqn\thhom{\Omega= \Omega_{K3} \wedge \th~,}
where $\Omega_{K3}$ is the holomorphic $(2,0)$-form on the
$K3$ base and $\th=(dz+\a)$ (with $z$ the fiber coordinate and $\a$
a connection one-form on $K3$ \foot{To be rigorous, $\alpha$ should be the pull-back of the
connection one-form on $K3$ to $X$.}) is defined to be a global
$(1,0)$-form. The hermitian $(1,1)$-form $J$ comes from the K\"ahler
form of the $K3$ and the metric on the torus bundle
\eqn\FYsol{J = e^{2\phi} J_{K3} + \frac{i}{2} \frac{A}{\tau_2}(dz+\a)\w (d\bz + {\bar \a})~,}
where the moduli of the torus are parametrized by the area $A$ and the
 complex structure $\tau=\tau_1 + i\, \tau_2$.   The dilaton $\phi$ is
  non-constant and depends on the base coordinates only. Supersymmetry demands that the
curvature of the torus bundle is of type $(2,0) \oplus (1,1)$
\eqn\tcurv{\om=\om_1+\tau\om_2=d \th = d \a \in H^{(2,0)}\oplus
H^{(1,1)}(K3) ~~~~{\rm with~~}  \om_1, \om_2 \in H^2(K3,\ZZ)~.}
Additionally, $\omega$ is required to be primitive with respect to the base
\eqn\prim {\om \w J_{K3} =0~.}
Turning on a $(2,0)$ component for $\omega$ reduces the ${\cal N}=2$
four-dimensional supersymmetry to ${\cal N}=1$, while a $(0,2)$
component breaks supersymmetry completely.\foot{A derivation of these
supersymmetry constraints can be found in the appendix of \bty.}

The solution also includes a gauge bundle with a gauge field
strength that satisfies the hermitian-Yang-Mills conditions
\hermitym. The field strength is related to the metric and curvature
two-form by the anomaly equation \anomcan. Integrating the anomaly
equation over the base $K3$ leads to the topological condition
\eqn\kfcon{ -\frac{1}{16\pi^2}\int_{K3}{\rm tr}\, F\w F - \frac{A}{\tau_2}\int_{K3}
\om\w {\bar \om}= 24 ~,} where ${\bar \om}= \om_1+{\bar \tau}\om_2$
is the complex conjugate of $\om$. This is the main sufficiency
condition to ensure that the anomaly equation, which for these
geometries can be interpreted as a non-linear second order partial
differential equation for $\phi$, can be solved \refs{\FY,\BBFTY}.

As for ordinary Calabi-Yau manifolds (see \vc), it was shown in
\becktwo\ that the net number of generations is determined by an
index which (for bundles with $c_1(F)=0$, such as the $SU(N)$
bundles we are considering) is related to the third Chern number of
the bundle \eqn\gen{N_{gen}={1\over 2}c_3(F)~.} More concretely, the
third Chern class of the bundle is \eqn\third{c_3(F)=-{i\over
48\pi^3}[2\,{\rm tr}\,(F\wedge F\wedge F)-3\,{\rm tr}\,(F\wedge
F)\wedge {\rm tr}\,F+{\rm tr}\,F\wedge {\rm tr}\,F\wedge {\rm
tr}\,F]~,} where the trace is taken over the generators of the
Lie-algebra. The generators are traceless for $SU(N)$ gauge groups,
so that trivially the first Chern class, $c_1(F)=0$. Thus for
$SU(N)$ bundles, the last two contributions to the above formula
vanish and integrating the first term gives the index of the
 Dirac operator \refs{\cylap, \kiyi}. If the spin connection is embedded into the gauge
  connection (which is not the case of interest here) this formula reduces to the more
  familiar expression $N_{gen}=\chi/2$, where $\chi$ is the Euler characteristic of the
internal manifold.  It was shown in \BBFTY\ that all stable gauge
bundles on $X$ that can satisfy the anomaly condition \anomcan\ are those obtained by
 lifting stable bundles on $K3$ to $X$.  From \third, it then becomes evident that the net number of generations vanishes.  Moreover, the existence of a non-vanishing vector field along the torus fiber implies that the Euler characteristic $\chi(X)=0$  \refs{\gp, \beck}.  
 
Some of the new torsional manifolds we construct in the next sections will have a non-vanishing Euler characteristic.  Though not given here, a more detailed understanding of the resolution of orbifold singularities of non-K"ahler geometries may reveal whether the resolution may allow gauge bundles with non-vanishing third Chern class.

\newsec{Orbifolding $T^2$-Bundles over $K3$}

New solutions can be constructed by orbifolding the $T^2$ bundle
over $K3$ geometry $X$ by a discrete symmetry group $\Gamma$ to
obtain $Y=X/\Gamma\,$.  We will discuss the characteristic features
of such orbifolds in this section and give explicit models in the
following sections.

When constructing orbifold models, we will in general only require that the
physical fields $(g,H_3,\phi, F)$ remain invariant under the orbifold
action $\Gamma$.  FSY geometries are supersymmetric and thus solve the supergravity
 equations of motion which are written in terms of the physical fields.  Thus, as long
  as the physical fields are invariant under $\Gamma$, the orbifold solution will
   also be a solution of the equations of motion.

However, the resulting orbifold solution will in general not be supersymmetric.
Preserving supersymmetry will require additionally that the pair $(J,\Om)$ is also
invariant under $\Gamma$.  As $J$ and $\Om$ are bilinears of the spinor $\eta$ that
generates ${\cal N}=1$ supersymmetry,
 {\it e.g.} $J_{mn}= i \eta^\dagger \gamma_{mn} \eta\,$ and $\Om_{mnp}=
 {\bar \eta}^\dagger \gamma_{mnp} \eta\,$, a nowhere vanishing $\eta$
  implies $(J, \Om)$ must be globally well-defined on $Y$ and hence must
  be invariant on $X$ under the action of $\Gamma$.

The relations between the physical fields and $(J,\Omega)$ are given
in \jrel-\prel.   From there, we see that $J$ must be invariant under
$\Gamma$, since it is directly related to the physical metric.  A central component
of $J$ is $J_{K3}$ which by uniqueness of the Calabi-Yau metric remains unchanged
under $\Gamma$ as long as the discrete symmetry action leaves invariant the K\"ahler class
 and the complex structure of the base $K3$.  $\Om$
however need not be invariant; the holomorphic $(3,0)$-form can transform by a
phase, $\Gamma: \Om\to \zeta\,\Om$ for $|\zeta |=1\,$, and still leave
$\p$ invariant in \prel\ and the complex structure of $X$ unchanged.  Thus, if $\zeta\neq 1$
then the resulting orbifold solution breaks all supersymmetry.

An element of the discrete symmetry group acting on $X$ can be thought of
as being composed of two components: $\rho=(\rho_1, \rho_2)$, where $\rho_1$
 acts on the $K3$ base while $\rho_2$ acts on the torus fiber.   It is useful
  to consider each action, $\rho_1$ and $\rho_2$, separately.  On the $K3$, finite
   group actions are standardly categorized by their actions on the holomorphic (2,0)-form.
      If an action leaves invariant $\Om^{2,0}$, {\it i.e.} $\rho_1 :
       \Omega^{2,0} \to \Omega^{2,0}\,$, then it is called a symplectic automorphism.
        If instead $\rho_1: \Omega^{2,0} \to \zeta \Omega$ for $\zeta\neq 1$, it is a
         called a non-symplectic automorphism.  In general, the action $\rho_1$ on $K3$
          will have a non-empty fixed locus set which may consist of points and/or curves.
           On the other hand, the symmetry group action on the torus is much more limited.
            For a group of order $N$, we can have either a shift $\rho_2: z \to z+c$ for
             some complex constant $c$ such that $Nc\sim z + n_1 + n_2\tau \,$, or a rotation,
             $\rho_2 : z \to \zeta\,  z$ with $\zeta^N =1$.

Because the torus bundle is generically non-trivial and does not have a zero section,
not any combination of $(\rho_1, \rho_2)$ will give a consistent symmetry
action on $X$.  Of importance, $X$ has a globally defined $(1,0)$-form $\th= dz + \alpha$
 which must have a well-defined transformation rule under $\Gamma$.  We require for
 any $\rho\in \Gamma\,$
\eqn\gamreq{\rho [\th] = \rho_2 [dz] + \rho_1 [\alpha]\,=\zeta\, \th~.}
We see for instance that a rotation on the torus fiber by itself can not be well-defined on
 $X$.  Consider the overlap region between two coordinate charts labelled by $A$ and $B$.
  The corresponding coordinates $z_A$ and $z_B$ and the local connection 1-form $\alpha_A$ and
  $\alpha_B$ are related as follows
\eqn\transfun{z_A = z_B + \varphi_{AB}~~~ {\rm mod~} 1\,, \tau~~, \qquad \alpha_A = \alpha_B -
 d \varphi_{AB}~,}
where $\varphi_{AB}$ is the transition function with dependence on the local base $K3$
coordinates.  Thus, a rotation acting on $z$ (compatible with the torus lattice
structure $\tau$) $\rho_2: z \to \zeta\, z\,$, only makes sense if there is a
 complementary action on the base $K3$ such that the transition function transforms
  as $\rho_1: \varphi_{AB} \to \zeta \,\varphi_{AB}\,$.  By \transfun, we see that
   the connection $\alpha$ must then transform similarly with the phase $\zeta$ which
    thus results in a well-defined transformation phase for $\th\,.$   Condition \gamreq\
     thus imposes a restriction on the allowable combination of $(\rho_1,\rho_2)$; and
     moreover, this restriction depends on the curvature $\om=d\alpha$.

Let us now consider the distinguishing features of the orbifold solution for different
types of symmetry group actions.  We shall order them by the action on the torus fiber.

For $\rho_2$ that involves only a torus shift, the action is
freely acting, {\it i.e.} without fixed points, and hence, the
resulting orbifold geometry is always smooth.  The amount of supersymmetry that is preserved
depends on the action on $\Om^{3,0}=\Om^{2,0}\w \th$.  But since $\th=dz + \alpha$ must be
 invariant under the torus shift (with $\alpha$
invariant under $\rho_1$ so that \gamreq\ is satisfied), the $\rho_1$ action
 on $\Om^{2,0}$ determines whether supersymmetry is preserved.  Thus, for symplectic
 $K3$ automorphisms, $Y=X/\Gamma$ will preserve the supersymmetry of $X$ while
  the non-symplectic automorphism action will break all supersymmetry.
  The base action will in general have fixed points and/or fixed curves.
  From the fiber bundle description, the complex structure of the torus will
   jump along the fixed point locus of the base.  Thus the torus bundle becomes a
    torus fibration.  Clearly, $Y$ is topologically distinct from $X$ as the first
     fundamental group of $Y$ now contains $\Gamma$.  Moreover, we note that the
      Betti numbers of $Y$ will generally differ from those of $X$ as the orbifolding
       by $\Gamma$ (with a non-trivial $\rho_1$) will project out non-invariant
        harmonic forms.  Nevertheless, the Euler characteristic and $c_3(F)$ will
        remain zero as the orbifold is freely acting.

For the case where the torus fiber action involves a rotation, there
should be an associated symmetry action on the base $K3$ by \gamreq.
Moreover, the discrete symmetry group $\Gamma$ will in general have
a fixed locus set which is invariant under
 $(\rho_1, \rho_2)\,$.  The resulting orbifold geometry $Y$ will then be
singular along the fixed points and/or fixed curves.  These
singularities will
 need to be resolved to obtain again a smooth solution.\foot{Singularity resolution can be thought of
as the geometrical analog of adding twisted sector states when constructing an
orbifold conformal field theory.}  Here, the task of resolving singularities
consists of two parts: resolving the manifold and smoothing out the physical
fields such that the supersymmetry conditions \balance-\anomcan\ (or the supergravity
equations of the motion) remain satisfied.

Resolving singular orbifolds $Y$ which are supersymmetric should
follow the
 standard local toric resolutions of singular Calabi-Yau manifolds (see {\it e.g.} \Greene)
by the requirement of a non-vanishing holomorphic $(3,0)$-form.  However,
unlike the Calabi-Yau case, where the vanishing of the first Chern class
is the sole obstruction to the existence of a Calabi-Yau metric on a K\"ahler manifold,
we do not presently know what are the obstructions or sufficient conditions for the existence
of a solution to the heterotic supersymmetry differential constraints of \balance-\anomcan.
Without this, we must explicitly demonstrate the existence of solutions on a manifold on a
case by case basis.  Thus, the resolution of singularities in the non-K\"ahler scenario
 is a challenging mathematical problem.  It requires constructing local models (such as \FTY) that smooth out the singularities and carefully gluing the local geometries into $Y$.

Nevertheless, having orbifold singularities may lead to new non-K\"ahler solutions
that might be phenomenologically appealing.  For instance, in resolving the singularities,
it may be possible to introduce local gauge bundles which have non-trivial $c_3(F)\,$, and hence non-zero number of generations.   In essence, one may try to satisfy the requirements of low-energy phenomenology by inserting appropriate local models into the compact geometry.  We shall however leave the subtleties of non-K\"ahler singularity resolutions for future work.

In the following sections, we will give concrete constructions of new
heterotic solutions from orbifolding $X$ by discrete symmetries.   This
requires identifying those $K3$ surfaces with a discrete symmetry group action.
We will consider three types of $K3$ surfaces that naturally contain discrete
symmetries: $K3$ surfaces of Kummer type, $K3$ surfaces
with a branched covering description, and more general algebraic ({\it i.e.} projective)
$K3$ surfaces with finite group actions.

\newsec{Orbifold Limit of Kummer $K3$ base}

In this section, we construct new orbifold geometries starting
from those FSY geometries with a Kummer $K3$ base.  A Kummer $K3$
 has a $T^4/\IZ_2$ orbifold limit with the involution ({\it i.e.} $\ZZ_2$ action) given by
\eqn\sigact{\sigma: (\za,\zb) \to (-\za,-\zb)~,}
where $(\za, \zb)$ are the complex coordinates of $T^4$.  The smooth Kummer
surface is obtained by blowing-up each of the $16$ singular fixed points of the
 involution $\sigma$ with a $\IP^1$.  Away from the singularities, the hermitian
  metric and holomorphic $(3,0)$-form take the simple form
\eqn\jokum{\eqalign{J& = \frac{i}{2}\,e^{2\p}(d\za\w d\bza + d\zb \w d\bzb) +
i\frac{A}{2}  \,\th \w \bth,\cr \Omega&=d\za\w d\zb\w \th,}}
where $\phi$ is the dilaton field and $\th=dz + \a$ denotes the one-form associated
to the twisted torus fiber.  For simplicity, we have set the complex structure of the
three covering-space tori to be $\tau=i$.  Thus for the base coordinates $(\za, \zb)$
 and fiber coordinate $z$, we identify $z_i\sim z_i + 1 \sim z_i+ i\,$, for $i=1,2$,
 and $z\sim z+1 \sim z+i$.  Given a torus twist curvature $\om$, we can then set the torus
  fiber area $A$ to satisfy the topological condition \kfcon.

We shall work mainly in the $T^4/\ZZ_2$ orbifold limit of the Kummer
 $K3$.\foot{Some earlier discussion on torsional orbifolds obtained via duality
 from Borcea fourfolds in M-theory appeared in \becktwo.}  This simplification will
 allow us to write down explicitly the discrete symmetry group $\Gamma$ acting on $X$
  and demonstrate clearly some of the characteristics of the new orbifold geometries
   discussed in section 3.  The discrete symmetry groups $\Gamma$ which we consider are
    all cyclic and thus are generated by a single element which we will denote by $\rho$.

\subsec{${\cal N}=1,2$ Supersymmetric Orbifolds}

Different types of orbifold actions will leave unbroken different
amounts of supersymmetry in four dimensions. We consider first those orbifold actions
 that leave invariant the holomorphic (3,0)-form and preserve the supersymmetry of the
  covering FSY geometry.   We will give four examples below and will point out the
  distinctive features of each.

\vskip 0.5cm

\noindent{\it Example 1: Freely acting with a shift action on the torus fiber}

We start with a simple model that has no fixed points, and therefore has a
vanishing Euler characteristic and a vanishing number of generations.
 Consider the $\ZZ_2$ action generated by
\eqn\rhoact{\r: (\za,\zb,z)\to (i\za,-i\zb, z+\frac{1}{2})~,} which
rotates the base in an $SU(2)$ invariant manner and shifts the torus
fiber. For the above discrete identifications, it is easy to see
that on the four-dimensional
 base, the $\ZZ_2$ action has four fixed
points located at
\eqn\bpts{ (\za,\zb)=\{ (0,0), (0, \frac{1+i}{2}), (\frac{1+i}{2}, 0),
 (\frac{1+i}{2},\frac{1+i}{2})\}~.}
But since the torus fiber is concurrently shifted by $z\to z+1/2\,$,
 the six-dimensional quotient manifold does not have any singularities and is thus smooth.

The above background is supersymmetric because, as follows from
\jokum, the pair $(J,\Om)$ is invariant under $\rho$ as long as
$\th=dz+\alpha$ is invariant. This can be easily satisfied by choosing a
curvature twist $\om =d \th\in H^{(2,0)}\oplus H^{(1,1)}(T^4/\ZZ_2,
\ZZ)$ that is primitive with respect to the base and invariant under
$\rho$.  For example, $\om$ can be any linear combination of $d\za \w d\zb$
 and $d\za \w d\bza - d\zb \w d\bzb$.  As shown in \bty, theories with a twist
 of type $(1,1)$ have ${\cal N}=2$ supersymmetry while a more general twist of type $(2,0)+(1,1)$
leads to a theory with ${\cal N}=1$ supersymmetry.  Similarly, for
the gauge field, we can choose $U(1)$ gauge field strengths
$F=F^{(1,1)}$ which are invariant under $\rho$.   This then ensures
that the three supersymmetry constraint equations \balance-\anomcan\
are invariant under $\r$. This background also naturally satisfies
the equations of motion since the physical fields $(g_{a\bar b}, H,
\phi, F_{a \bar b})$ are invariant under the orbifold action.

As a fiber space, this smooth quotient geometry should be considered as a
 $T^2$ {\it fibration} rather than a $T^2$  {\it bundle} as the complex
  structure of the $T^2$ fiber jumps at the four singular points \bpts\ of
   the base identified by the involution action $\rho$.
At the fixed points, the complex structure jumps to $\tau=2i$ with
$z\sim z+ \frac{1}{2} \sim z + i\,$.

Finally, we note also that the $\ZZ_2$ action \rhoact\ preserves not only
the holomorphic $(3,0)$-form but also the holomorphic $(2,0)$-form
of the base $K3$. Its action on the base $K3$ is thus an example of a symplectic
automorphism, a discrete symmetry group that preserve $\Om^{2,0}$.

\vskip 0.5cm

\noindent{\it Example 2: Freely acting with a reflection on the torus fiber}

It is also possible to consider a freely acting involution which
involves
 a non-symplectic automorphism, one that acts non-trivially on the holomorphic
 $(2,0)$-form on the base $K3$.  One such involution is generated by
 \eqn\rhoenr{\r: (\za,\zb,z) \to (-\za + \frac{1}{2}, \zb +\frac{1}{2}, -z)~.}
$\rho$ here acts freely on the base coordinates $(\za,\zb)$, and has been called
an Enriques involution \hit\ since the base holomorphic $(2,0)$-form is mapped
to the minus of itself \eqn\ei{\r: \Om_{K3}\to -\Om_{K3}.}  The holomorphic
$(3,0)$-form however remains invariant when we take into account the
compensating reflection action, $z\to -z$, on the fiber coordinate.

Precisely the involution in \rhoenr\ acting on the product space
$K3\times T^2$, was analyzed in \refs{\vawi,\hls} in the context of
type IIA/heterotic string duality. That such a quotient can also be
applied to the FSY geometry with a non-trivial principal
$T^2=S^1\times S^1$ bundle
 with no zero section may at first seem surprising.  But as explained earlier, as
  long as we carefully choose a torus twist that gets reflected along with the
   fiber coordinate such that the one-form $\th = dz + \a$ has a well-defined action,
   then the $\ZZ_2$ action \rhoenr\ consisting of a fiber reflection coupled with an
   involution on the base is well-defined.  Let us verify this explicitly for this example.

First, a consistent torus twist for the $\ZZ_2$ action \rhoenr\  is
\eqn\rhoenra{\th = dz + A_1 (\za -\bza)d\zb  + A_2 (\zb - \bzb) d\za~,}
where $A_1$ and $A_2$, are Gaussian integers (complex numbers with integral real and
imaginary parts).  Note that with this one-form, the torus twist curvature $\om = d\th$
contains both a $(1,1)$ and a $(2,0)$ part so the covering FSY geometry and also
the orbifold geometry preserve only ${\cal N}=1$ supersymmetry.

With the one-form \rhoenra, let us demonstrate explicitly the consistency of the quotient action.
We will work in the covering space which is a $T^2$ bundle over a $T^4$
base.  The metric on this space (neglecting the warp factor and the torus
area $A$ which do not a play a role here) takes the form
\eqn\amet{ds^2 = |d\za|^2 + |d\zb|^2 + |dz+ A_1 (\zb -\bzb)d\za  +
A_2 (\za - \bza) d\zb|^2~.}
In order for the metric to be well-defined, the local complex coordinates must
have the periodicity
\eqn\period{\eqalign{\za&\sim \za+a~ ,\cr \zb &\sim \zb +b~,\cr z~&\sim z+ c - A_1
(b -{\bar b}) \za - A_2 (a - {\bar a}) \zb~,}}
where $a,b,c$ are arbitrary Gaussian numbers.  These periodic boundary conditions
define the transition functions on the manifold.  The quotient action \rhoenr\
acts on the periodicity \period\ and results in
\eqn\rperiod{\eqalign{
-\za  &\sim - \za + a = -(\za + a')  ~,\cr
\zb  &\sim \zb + b  = \zb + b' ~,\cr
-z\,\,&\sim -z+ c + A_1 (b -{\bar b}) (\za - \frac{1}{2})   - A_2 (a - {\bar a})
 (\zb + \frac{1}{2}) \cr & = -(z+ c' - A_1 (b' -{\bar b'}) \za - A_2 (a' - {\bar a'}) \zb)  }}
where we have defined
\eqn\cdef{(a', b', c')= (- a ,  b , -c + i A_1 {\rm Im}(b) + iA_2 {\rm Im}(a))~.}
Since the constants $a,b,c$ are arbitrary Gaussian numbers, the
redefinition of \cdef\ is inconsequential. Therefore, from  \rperiod,
 we see that the quotient action preserves the periodicity \period\ of the
  covering space. This implies that the quotient is well-defined.

\vskip 0.5cm

\noindent{\it Example 3: An involution with fixed curves}

From the previous example, we have seen how a discrete symmetry
action can involve a reflection, a simple example of a rotation
action on the fiber. Coupled with the discrete symmetry action on
the base, the discrete group action with fiber rotation action will
generally have fixed points and/or fixed curves.

Consider the $\ZZ_2$ action generated by
 \eqn\rhonss{\rho: (\za,\zb, z)\to (i\za, i\zb, -z)~.}
This is an involution since $(\rho)^2$ is equivalent to $\sigma$, the $\ZZ_2$
action appearing in \sigact. This involution again involves a reflection on the
 $T^2$ bundle. As above we need to require that the curvature twist transforms
  $\om \to -\om$ under
$\rho$.  For this we must have $\om= d\theta \sim d\za\w d\zb$ and
this will
 ensure that the one-form $\th$ is well-defined under the involution.

The quotient manifold however will have fixed curves.  On the $T^6$
covering space, the $\sigma$ and $\rho$ actions result in 16 fixed
points - four fixed points on the base times four fixed points on
the fiber.  The four base fixed points (those of \bpts) coincide
with the singularities of the $T^4/\ZZ_2$ orbifold and are resolved
by $\PP^1$'s when the base $K3$ orbifold is resolved. These
$\PP^1$'s however are invariant under $\rho$ and result in $16$
fixed curves.

We can explicitly see this for instance in the resolution of the
point $(\za,\zb,z)=(0,0,0)$.  Locally, this is equivalent to
resolving the origin of $C^3/(\IZ_2 \times \IZ_2)$ with the quotient
 generated by $\sigma $ and $\rho$.  This local orbifold can be
minimally resolved following the toric resolution methods of
\refs{\MOP, \RoanYau}. We resolve the singularity in two steps.   We
first resolve the singularity at the origin of $C^2/\{1,\sigma\}~ \times
C$ and then proceed to quotient by $\rho$ and resolve again.

For $C^2/\{1,\sigma\}\, \times C$, the origin is resolved by a $\PP^1$.
Following e.g. \MOP, the resolution is covered by two coordinate
charts: \eqn\schart{\eqalign{U_1:&\quad (\frac{\zb}{\za},\za^2,z)~,
\cr U_2:& \quad (\frac{\za}{\zb},\zb^2,z)~.}}
Here, the first entry
is the coordinate of the $\PP^1\,$: $\xi=\frac{\zb}{\za}$ in $U_1$
and $\xi'=1/\xi$ in $U_2$.  We can proceed to apply the $\rho$ action
\rhonss\ on the two coordinate charts:
\eqn\rchart{\eqalign{U_1:\qquad
\rho:&~\left(\frac{\zb}{\za},\za^2,z\right)\to\left(\frac{\zb}{\za},-\za^2,-z\right)~,\cr
U_2:\qquad
\rho:&~\left(\frac{\za}{\zb},\zb^2,z\right)\to\left(\frac{\za}{\zb},-\zb^2,-z\right)~.}}
We see that $\rho$ leaves the $\PP^1$ coordinate $\xi$ (and $\xi'$)
invariant and acts only on the two transverse coordinates. This
shows that the $\PP^1$ curve is invariant under the $\rho$ action.
To resolve the two transverse coordinates, which is another
$C^2/\{1,\sigma\}$ singularity, we can repeat the resolution of \schart\
on $U_1$ and $U_2$ charts separately. The total resolution is
therefore described by four coordinate charts: $U_{11}, U_{12},
U_{21}, U_{22}$.
\eqn\srchart{\eqalign{U_{11}:\quad\left(\frac{\zb}{\za},
\frac{z}{\za^2}, \za^4\right)\qquad\qquad &U_{21}:\quad
\left(\frac{\za}{\zb},\frac{z}{\zb^2},\zb^4\right)\cr
U_{12}:\quad\left(\frac{\zb}{\za},\frac{\za^2}{z},z^2\right)\qquad\qquad&U_{22}:\quad
\left(\frac{\za}{\zb},\frac{\zb^2}{z},z^2\right) }}
On every point of the $\PP^1$ curve, we have added another $\PP^1$.
Thus the divisor is a ruled surface, and by the fibration structure, it is a
Hirzebruch $F_2$ surface which has Euler characteristic
$\chi(F_2)=4$. After resolving all $16$ $\PP^1$'s$\,$, the Euler
characteristic of the resolved orbifold $\widehat Y$ is
\eqn\chixx{\chi(\widehat Y)= [\chi(X)-16 \,\chi(\PP^1)]/2 + 16\, \chi(F_2)
= -16 + 64 = 48~.}

We have thus described how to resolve the singular orbifold into a smooth
manifold with non-zero Euler characteristic.  If we are only interested in
Calabi-Yau solutions, then this is sufficient to describe the solution as
 Yau's theorem implies the existence of a Ricci-flat metric.  However, without
  a corresponding theorem for the existence of non-K\"ahler heterotic solutions,
   we have to  demonstrate explicitly that a non-K\"ahler balanced metric exists for this
   resolved manifold.  This may be done by constructing non-compact solutions that
   locally resolved the singularities.  One would then need to cut out the singularities
   and carefully glue in these local solutions into the manifold.

\vskip 0.5cm

\noindent{\it Example 4: A $\ZZ_4$ quotient with fixed points and fixed curves singularities}

In general,  the generic orbifold model will have both fixed points and fixed
curves singularities which will need to be resolved.   Let us give an example.

Consider the $\ZZ_4$ action generated by \eqn\rhofp{\r: (\za,\zb,
z)\to (i\za, -\zb, iz)~.} Though very similar to the previous
example, this action does not square to the $\ZZ_2$ identification
\sigact\ of the $K3$.  Thus it is a $\ZZ_4$ action. In order for
$\th$ to be well-defined, we are constrained to require that the
torus curvature twist transforms as $\om \to i\, \om$, which implies
that $\om \sim d\bza \w d\zb$.

The action \rhofp\ has eight fixed points on the base given by
\eqn\bptfp{(\za,\zb)=\{(0,0),(0,\frac{1}{2}),
(0,\frac{i}{2}),(0,\frac{1+i}{2})
(\frac{1+i}{2},0),(\frac{1+i}{2},\frac{1}{2}),(\frac{1+i}{2},\frac{i}{2}),
(\frac{1+i}{2},\frac{1+i}{2})\},} and two fixed points on the fiber
at $z=\{0,\frac{1+i}{2}\}$. The base fixed points again coincide with
those of the orbifold $T^4/\ZZ_2$ and are resolved by $\PP^1$'s.
However, these $\PP^1$'s are not invariant under $\rho$ as defined
in \rhofp. We can see this from the resolution of the point
$(0,0,0)$. The action of  $\rho$ on the two charts of
$\PP^1$ is
\eqn\rfchart{\eqalign{U_1:\qquad
\rho:&~\left(\frac{\zb}{\za},\za^2,z\right)\to\left(i\frac{\zb}{\za},-\za^2,iz\right)~,\cr
U_2:\qquad
\rho:&~\left(\frac{\za}{\zb},\zb^2,z\right)\to\left(-i\frac{\za}{\zb},\zb^2,iz\right)~.}}
We see that $\rho$ acts differently on the two charts.  But in each
case, the resolved $\PP^1$, denoted by the first coordinate, is also
rotated by the $\rho$ action. So considering the base and fiber
together, we have here 16 true fixed points.

There are also fixed curves in this model. They arise from points
which are fixed under $\r^2$ but not $\r$.  The $\r^2$ action,
$\r^2: (\za,\zb,z)\to(-\za,\zb,-z)$, has fixed curves at
\eqn\bptfc{\eqalign{(\za,z)&={\Big\{}(0,\frac{1}{2}),(\frac{1+i}{2},\frac{1}{2}),(\frac{1}{2},0),
(\frac{1}{2},\frac{1}{2})
,(\frac{1}{2},\frac{i}{2}),(\frac{1}{2},\frac{1+i}{2}), \cr & \qquad
 (0,\frac{i}{2}),(\frac{1+i}{2},\frac{i}{2}),
(\frac{i}{2},0),(\frac{i}{2},\frac{i}{2}),(\frac{i}{2},\frac{1}{2}),(\frac{i}{2},\frac{1+i}{2}){\Big
\}}~. } }
These curves are not invariant under $\rho$; in fact, the
curves on the first line of \bptfc\ are identified with those on the
second line in the local coordinates $(\za,z)$.  For $\zb\neq 0 ,
\frac{1}{2},\frac{i}{2},\frac{1+i}{2}$ which are the fixed points of
$\zb$ under $\r$, the $12$ curves persist.  At the $\zb$ fixed
points, the identification reduces to $6$ distinct curves.

%%%%%%%%%%%%%%%%%%%%%%%%%%%%%%%%%%%%%%%

\subsec{Non-Supersymmetric Orbifolds}

Non-supersymmetric models can easily be constructed similar to the examples given above.
The orbifold action now is required to act on the holomorphic three-form non-trivially.
 Consider the  $\ZZ_2$ action generated by
\eqn\rhons{\rho: (\za,\zb,z)\to (i\za,i\zb,z+1/2)~.} Here, $\r$
differs from the action in \rhoact\ by a minus sign in the action on
$\zb$.  As a result, the base holomorphic two-form
$\Om_{T^4/\ZZ_2}=d\za \w d\zb$ picks up a minus sign and the
holomorphic three-form transforms non-trivially, {\it i.e.} $\Om \to
-\Om\,$ and is thus projected out. Therefore, this type of solution
is not supersymmetric.

The torus curvature $\om=d\th$ and $U(1)$ field strength $F$ are required to be primitive
 and invariant under $\rho$.  A basis of such $(1,1)$ forms
 is $\{d\za\w d\bzb,d\zb\w d\bza, d\za \w d\bza - d\zb \w d\bzb\}$.  All
 the physical fields $(g, H_3, \p, F)$ of the model can remain invariant under $\r$.
Thus, this gives a smooth non-supersymmetric solution of the
supergravity equations of motion.

\newsec{Orbifolds from Branched Covered $K3$ Base}

We are interested in complex manifolds whose first Chern class $c_1(M)$ is zero.
 When the canonical bundle is non-trivial, {\it i.e.
$c_1(M)\neq 0$}, it is sometimes possible to eliminate the first
Chern class by taking $n$ copies ({\it aka} covers) of the manifold
$M$ and glue them together at a divisor, a codimension one
hypersurface.  More precisely, this new manifold $M'$, which is
described as an $n$-fold cover of $M$ branched over a divisor $D$,
can have trivial canonical bundle.\foot{For examples of Calabi-Yau
three-fold with a branched covering description, see \refs{\yau, \StrWi, \Hubsch}.}
The new Chern class is given by
\eqn\bchern{c_1(M')=n[c_1(M)-c_1(D)]+ c_1(D)=n\, c_1(M) - (n-1)c_1(D).}
A Calabi-Yau manifold with an $n$-fold branched covering
description clearly has a manifest $\IZ_n$ discrete symmetry acting on the $N$ identical covers.
In this section, we construct orbifold solutions starting from FSY geometries
containing a $K3$ base with a branched covering description.   Below, we will
 make use of two examples of branched covered $K3$ surfaces: a triple cover
of $\IP^1 \times \IP^1$ branched over a genus four curve and a double
cover of $\PP^2$ branched over a genus ten curve.

Before jumping into the details of branched covered $K3$ surfaces,
let us first illustrate the basic idea of using branched covering
for the simpler complex dimension one case where the first Chern
number agrees with the Euler
 characteristic $C_1(M)=\chi(M)$.  Let $M=\PP^1=S^2$.~~Being in two real
  dimensions, $\chi_1(M')=0$ means $M'=T^2$ for compact manifolds.  Since
  $\chi_1(M)=2$ and $\chi_1(D)$ equals the number of branched points, we see from \bchern\ that one
can construct a torus as a two-fold cover of $\PP^1$ branched over
four points or as a three-fold cover of $\PP^1$ branched over three
points.   Reversing the construction, we can start with the double
covered torus $M'=T^2$ and quotient by an involution to obtain a
$\PP^1$ with four fixed points, or with the triple covered torus and
quotient by a $\ZZ_3$ action to obtain a $\PP^1$ with three fixed
points.  Such natural quotienting can be
 applied similarly to branched covered $K3$ surfaces.

\subsec{$\PP^1\times \PP^1$ Base Solution}

We take the base $K3$ surface to be a triple cover of $\PP^1 \times \PP^1$
branched over a sextic curve.  As described below, this $K3$ is a
complete intersection of a quadric and a cubic equation in $\PP^4$.
Using this $K3$ to construct a FSY geometry, we can quotient by the natural $\ZZ_3$
 action and obtain an orbifold non-K\"ahler heterotic solution that is an elliptic
  fibration over a $\PP^1\times \PP^1$ base.

To describe the $K3$,  let $\{\zo,\za,\zb,\zc,\zd\}$ be the homogeneous coordinates of
$\PP^4$.  The $K3$ hypersurface in $\PP^4$ is defined by the
following two equations \eqn\fa{f^1=\zo\zc - \za\zb = 0~,}
\eqn\fb{f^2=g(\zo, \za,\zb,\zc) + \zd^3 =0~,} where $g$ is a degree
three homogeneous polynomial in $z_i$.  The first equation
enforces the standard embedding of $\PP^1\times\PP^1 \to \PP^3\subset
\PP^4$ by the mapping \eqn\ppt{\matrix{\PP^1 &\times& \PP^1  &\to
&\PP^3 \cr \{\xo, \xa\} &\times& \{\yo, \ya\} &  &
\{\zo,\za,\zb,\zc\}=\{\xo\yo, \xa\yo, \xo\ya,\xa\ya\}.}} The second
equation exhibits the three-fold covering.  For each generic point
on $\PP^1 \times \PP^1$, there are three different values of $\zd$
that satisfy \fa.  Alternatively, the equations above are invariant
under the $\ZZ_3$ action generated by
\eqn\ztact{\rho: \zd \to \zeta \zd~,  \qquad {\rm
where} \qquad \zeta =e^{2\pi i /3}~.}
In particular the special point $\zd =0$ is invariant under this action and
 here we have the branched curve specified by
\eqn\bcurve{g(\zo,\za,\zb,\zc)=0~.}
This is a cubic equation in terms of the $z_i$ variables and a sextic
equation in terms of the natural homogeneous variables $x_i, y_i$ on
$\PP^1\times \PP^1$.  To ensure the hypersurface is smooth, it is
sufficient to require that the normal bundle to the hypersurface
does not vanish. That is
\eqn\smcond{df^1\w df^2 = (\zc dz_0 - \zb dz_1 - \za d\zb +
\zo d\zc) \w (\sum_{i=0}^3 \pa_ig \,dz_i + 3 \zd^2 d\zd)\neq 0~,}
for any points on the hypersurface.  This is a constraint for $g$.  For
 instance, it can be checked that a hypersurface defined by $g=\zo^3+\za^3+\zb^3+\zc^3$
 is singular for example at $(\zo,\za,\zb,\zc)=(1,-1,-1,1)$ while $g=\zo^3+\za^3+\zb^3+2\,\zc^3$, the
 example we shall consider below, is everywhere smooth.

We can write down explicitly the holomorphic two-form for the $K3$
hypersurface.  (For reference, see {\it e.g.} \GSW\ Section 15.4.) In the
local chart where $\zo\neq 0$, we define the affine coordinates $Z_i=
z_i/z_0$ for $i=0,\ldots, 3$.  In these coordinates the constraint
polynomials become
\eqn\Fa{f^1=Z_3-Z_1Z_2=0~,} \eqn\Fb{f^2=g(1,\Za,\Zb,\Zc)+\Zd^3 =0~.}
The holomorphic two-form in terms of $d\Za \w d\Zb$ is then
\eqn\holt{\Om^{2,0} = d\Za \w d\Zb /
({\rm det} M)= \frac{d\Za\w d\Zb}{3\Zd^2}~,}
where the $2\times 2$ matrix $M_{ij}=\pa f^i/\pa Z_{2+j}$ consists of the partial
derivatives of the two other coordinates $(\Zc,\Zd)$.  The choice of
the $(\Za,\Zb)$ of course is arbitrary and we could have used any other
two coordinates.  For instance, in terms of $d\Zc\w d\Zd$, we have
\eqn\holtt{\Om^{2,0}= \frac{d\Zc \w d\Zd}{\Za\,\pa_1g - \Zb\,\pa_2 g}~.}
When \smcond\ is satisfied for a smooth $K3$ hypersurface,
the different det$M$ for different choices of coordinates never
simultaneously vanish. This must be the case since $\Om^{2,0}$ is no-where vanishing.
Moreover, of importance for our
construction, under the $\ZZ_3$ action, $\rho: \Zd\to \zeta \Zd$, the
holomorphic two-form is not invariant and in fact transforms as
$\Om^{2,0}\to \zeta\, \Om^{2,0}$. This is clearly evident in \holt\ and \holtt.
In contrast, the Calabi-Yau metric or the hermitian form $J_{K3}$ is invariant under $\ZZ_3$.

We now consider the non-singular example
with the $K3$ hypersurface defined by
\eqn\defg{\eqalign{f^1&= g(\zo,\za,\zb,\zc) + \zd^4
=\zo^3+\za^3+\zb^3+2\zc^3 + \zd^3,\cr &=(\xo^3 + \xa^3) (\yo^3 +
\ya^3) + \xa^3\ya^3 + \zd^3 = 0~,}}
where we have substituted in the coordinates of the two $\PP^1$'s, $x_i$ and $y_i$.
 The branched curve $C$ is located at $g=\zd=0$.  Assuming $\zo=\xo\yo\neq 0$, the
equation of the branched curve can be written as
\eqn\bcurve{(1+X^3)(1+Y^3)+X^3Y^3 =-\Zd^3 = 0~, } where $X=\xa/\xo$
and $Y=\ya/\yo$.  (Note from the relation $\chi(K3)= 3[\chi(\PP^1\times\PP^1) - \chi(C)] +
 \chi(C)$, we see that branched curve has genus $g=4$.)  Two distinguished curves in this
  model are obviously
the curves on each of the $\PP^1$.  These follow from \bcurve\ by
taking either $X$ or $Y$ to be fixed \eqn\ppcurve{\eqalign{
C_\Xo:\qquad& -\Zd^3 = 1 + Y^3 + \Xo^3(1+2Y^3)~, \cr C_\Yo:\qquad&
-\Zd^3 = 1 + X^3 + \Yo^3(1+2X^3)~, }} where $\Xo, \Yo \in {\bf C}~$
are complex constants. Note that the $C_\Xo$ and $C_\Yo$ curves are holomorphically
embedded into the $K3$ surface.  The two classes are also
homologously inequivalent.  The class of $C_{\Xo}$ curves do not
self-intersect while they intersect three times with the $C_{\Yo}$
curves.

Given the above $K3$ surface, we can build a FSY geometry.  We twist
the $T^2$ by a curvature $\om= \om_1 - \om_2$, where $\om_1$ and
$\om_2$ are the two forms dual to the curves $C_\Xo$ and $C_\Yo$.
The explicit expressions for these forms can be obtained by
Poincar\'e duality though we will not need them.  Notice that $\om = \om_1 - \om_2$
is primitive with respect to the K\"ahler form on $\PP^1 \times \PP^1$
and also $J_{K3}$.  Moreover, since the curves $C_{\Xo}$ and $C_{\Yo}$ are
 holomorphically embedded,
$\om \in H^{1,1}(K3)\cap H^2(K3, \ZZ)$. $\om$ can also be the curvature form of
any $U(1)$ gauge bundles that one wish to turn on.

This construction provides an explicit algebraic description of the covering FSY geometry.  We can now orbifold by a $\ZZ_3$ action to obtain a new solution.  Again, we can take the
fiber torus to have square periodicities $z \sim z+1 \sim
z+i$.  The $\ZZ_3$ action acting on the six-dimensional geometry is
 generated by
 \eqn\zact{\rho:\{z_0,\za.\zb,\zc,\zd,z\} \to \{\z_0,\za,\zb,\zc,\zeta \zd , z+ 1/3\}}
where $\zeta$ is a third root of unity, {\it i.e.} $\zeta^3=1$.  $\rho$
acting on the base reduces the triple cover to just a single copy of
$\PP^1 \times \PP^1$.  Without the $T^2$ bundle, we have a singular
branched curve that is invariant under $\rho$.   The shift action on
$T^2$ removes the singular identification, however, it also increases
the complex structure $\tau$ of the $T^2$ fiber along the branched curve.
The torus complex structure thus jumps along the branched curve.  But
nevertheless, the resulting geometry is smooth since there are no
fixed points.

One might worry about the $\ZZ_3$ action on the curvature $\om$.
But since $\om$ is dual to the curves, $C_1$ and $C_2$, which are
invariant under $\rho$, $\om$ must also be invariant under $\rho$.

The FSY geometry with the three-fold cover $K3$ surface is thus invariant under $\rho$.
 Since the holomorphic three-form $\Om = \Om^{2,0} \w \th$ transforms
 non-trivially $\rho: \Om \to \zeta\, \Om$, the smooth orbifold solution
 with an elliptic fibration over $\PP^1\times \PP^1$ is non-supersymmetric.

\subsec{$\PP^2$ Base Solution}

We can take the $K3$ surface to be a double cover over a $\PP^2$
base branched over a sextic curve.  The $K3$ surface is defined as a
degree six hypersurface in $W\PP^3_{1113}$.  With weighted
homogeneous coordinates $(\zo,\za,\zb,\zc)$, we can take the
hypersurface to be \eqn\ppbh{\zo^6 + \za^6 + \zb^6 + \zc^2 =0~.}
Notice that $(\zo,\za,\zb)$ defines a $\PP^2$.  And for each point
on $\PP^2$, there are two values of $\zc$ that satisfy \ppbh, except
on the degree six (genus ten) branched curve
\eqn\sbran{g(\zo,\za,\zb) = \zo^6 + \za^6 + \zb^6 = 0~.}

Similar to the $\PP^1\times\PP^1$ case, the $\Om^{2,0}$ form on the
base is not invariant under the $\ZZ_2$ action $\zc \to -\zc$.
However, we can preserve the holomorphic three-form by considering
the $\ZZ_2$ quotient generated by \eqn\Rquotient{\rho: (\zc,z) \to
(-\zc,-z).} Taking the twist curvature $\om \sim\Om^{2,0}$, ensures
that $\th \to -\th$ under $\rho$.  $\Om=\Om^{2,0}\w \th$ is
therefore invariant under the $\ZZ_2$ action.   The degree six
branched curve defined by \sbran\ is however singular. Thus, this
construction gives a singular ${\cal N}=1$ supersymmetric orbifold
solution.   As before, we leave the resolution of these singular
curves for future work.

%%%%%%%%%%%%%%%%%%%%%%%%%%%%%%%%%%%%%%%%%%%%%

\newsec{Algebraic $K3$ Surfaces with Finite Group Action}

In the previous section, we have seen two examples of $K3$ surfaces with a branched
covering description and with it a natural discrete symmetry group action.  In both
cases, the discrete symmetries are non-symplectic automorphisms
which by definition act non-trivially on the holomorphic two-form
$\Om^{2,0}$ of the $K3$.  More generally, $K3$ surfaces with non-symplectic
symmetries are algebraic and have been classified in \nika\ for
the $\ZZ_2$ case and  \refs{\arts,\taki} for the $\ZZ_3$ case.  These $K3$ surfaces can all be used to construct orbifold FSY geometries.  The general construction is  similar to the constructions given in the previous sections.  The main difference being the determination of the torus curvature 2-form $\om$ with the desired transformation property under the $K3$ discrete group action.  

Below we shall show how to go about explicitly writing down the torus twist for algebraic $K3$ surfaces, and in so doing, construct non-supersymmetric and singular supersymmetric orbifold geometries.  With algebraic $K3$ surfaces, our description becomes effectively purely algebraic.  We explain this in the context of a special class of algebraic $K3$ with a $\ZZ_3$ discrete symmetry group that has only three fixed points and no fixed curves \refs{\arts,\taki}.\foot{It follows from the holomorphic Lefschetz formula that any $K3$ surface that has a $\ZZ_3$ discrete symmetry with no fixed curves must have precisely three fixed points.}  This unique class of $K3$ is of particular interest since the singular orbifolds that are constructed from them will only have fixed points and therefore should be easier to resolve.  Though we focus on this special class of $K3$ surfaces, the orbifold construction we give below should be applicable to other algebraic $K3$ surfaces with a non-symplectic automorphism.

The class of algebraic $K3$ surfaces that we are interested in can be described, similar to the three-fold branched cover over $\PP^1\times\PP^1$, as an
intersection of a quadric and a cubic hypersurface in $\PP^4$.
Again, let $\{\zo,\za,\zb,\zc,\zd\}$ be the homogeneous coordinates
of $\PP^4$. The class of $K3$ hypersurface $S$ in $\PP^4$ is given by the
following two equations \arts 
\eqn\fat{f^1=f_2(\zo,\za)+ b_1\zb\zc +
b_2\zb\zd=0~,} \eqn\fbt{f^2=f_3(\zo,\za) + b_3\zb^3 + g_3(\zc,\zd) +
\zb f_1(\zo,\za) g_1(\zc,\zd)=0~,} where $f_n, g_n$ are homogeneous
polynomials of degree $n$ and $b_i$ are non-zero complex constants.
Notice that the quadric and cubic equations are invariant under the
following discrete action\foot{To simplify notation, we will in this section use $\rho$ to denote the generator of the discrete symmetry action acting either only on the $K3$ base, only on the torus fiber, or on the entire six-dimensional geometry $X$.  The object of the $\rho$ action should be clear from the context.} $\rho$ acting on $S$  
 \eqn\zthree{\rho: (\zo,\za,\zb,\zc,\zd)\to
(\zeta^2 \zo,\zeta^2\za,\zeta\zb,\zc,\zd)~.} 
The solutions of the two hypersurface equations with $(\zo,\za,\zb)=(0,0,0)$ give the three fixed points which solve $g_3(\zc,\zd)=0\,$.  

For instance, we can consider the $K3$ hypersurface defined by the homogeneous equations
\eqn\fatd{\eqalign{f^1&= \zo^2 + \za^2+ \zb (\zc + \zd)
\cr f^2&=\za^3 + \zb^3 + \zc^3 - \zd^3 }~.}
It can be checked that this $K3$ hypersurface is smooth with $df^1\w df^2 \neq 0$ on
the hypersurface of \fatd.  In the coordinate chart $\zc\neq 0$, we can
use the affine coordinates $Z_i=z_i/z_3$ for $i=0,\ldots,4$.  Then the three fixed
points of \zthree\ are located at
$(Z_3,Z_4)=\{(1,1),(1,\zeta),(1,\zeta^2)\}$.  Similar to \holt, the
holomorphic two-form can be written locally for example as
\eqn\omfat{\Om^{2,0} = \frac{dZ_0 \w dZ_1}{-3(Z_2^3 + Z_4^2)}~,} 
which shows explicitly that the holomorphic two-form transforms non-trivially under the $\ZZ_3$ action as 
\eqn\rhomalg{\rho: \Om^{2,0} \to \zeta \Om^{2,0}~,}
as expected for an non-symplectic automorphism.  

To construct orbifold FSY geometries, we need to define the torus curvature twist $\om=\om_1 + \tau \om_2$.  The conditions on $\om$, \tcurv\ and \prim, are that it is an element of $H^{2,0}(S)\oplus H^{1,1}(S)$ with $\om_1,\om_2 \in H^2(S,\ZZ)$ and that it is primitive.  

For the non-supersymmetric orbifold group action that contains a shift on the torus fiber, 
\eqn\zths{\rho: (\zo,\za,\zb,\zc,\zd,z)\to (\zeta^2\zo,\zeta^2\za,\zeta\zb,\zc,\zd,z+1/3)~.} 
the torus curvature $\om$ must be invariant under $\rho$.  Since $\Om^{2,0}$ is not invariant, we have $\om_1, \om_2$ can only be $(1,1)$  and are elements of the Picard lattice $N=H^{1,1}(S) \cap H^2(S,\ZZ)$.  In particular, they must be elements of the sublattice $N(\rho) \subset N$ that is invariant under $\rho$.  We now show that the $N(\rho)$ is a non-empty lattice and in fact has rank $\rk N(\rho) = 8$.  This can be derived using the Lefschetz fixed point formula which relates the Euler characteristic of the fixed point set
$S_\rho$ with the transformation properties of the de Rham
cohomology under $\rho$.  We denote the complement lattice $\Nkc$ of $N(\rho)$ in $H^2(S,\ZZ)$.  $\Nkc$ is not invariant under $\rho$ and let us assume it has rank $\rk \Nkc = 2m$.  Since the second Betti number $b_2=22$, we have $\rk \Nk= 22-2m$.  The Lefschetz fixed point formula then gives
\eqn\Lfp{\eqalign{3=\chi(S_\rho) &= \sum_{k=0}^4\Tr(\rho|_{H^k
(S,\ZZ)})\cr & = 1 + \rk \Nk + m (\zeta + \zeta^2) + 1 \cr & = 24-
3m }} 
where we have used the fact that $1+\zeta + \zeta^2 =0$.  This
implies that $m=7$ and thus $\rk \Nk= 8$ and $\rk \Nkc = 14$.  Thus, we need to choose primitive $\om_1,\om_2 \in N(\rho)$ for the torus curvature twist.  (The explicit lattice $N(\rho)$ is given in Eq. (6.11) below.)

For constructing supersymmetric orbifolds, the orbifold action must act on the torus fiber by a rotation 
\eqn\ktrn{\rho: \quad\matrix{ \th &\to& \z^2 \th\cr
\om &\to& \z^2 \om}}
so that the holomorphic three-form remains invariant.  In order for the torus lattice structure to be compatible with this $Z_3$ rotation, we must set the torus complex structure $\tau=\zeta$.   The torus curvature two-form $\om= \om_1 + \zeta \om_2$ is now required to have the transformation property $\rho: \om \to \zeta^2 \om$.  Thus, $\om_1, \om_2 \in \Nkc$ and because $\Om^{2,0}$ transforms as in \rhomalg, we must again have $\om \in H^{1,1}(S)$.  Note that elements in $\Nkc$ are real and generally, in addition to $(1,1)$ components, also  have $(2,0)$ and $(0,2)$ components.   Thus to ensure that $\om$ is purely $(1,1)$, we need to determine explicitly the complex structure of $S$.   To do so will require some knowledge of the lattice $L$ of the second integral cohomology, $H^2(S,\ZZ)$, which we now explain.  (For more details on $L$, see for example \Morris.)

$L$ is a self-dual unimodular lattice. For elements $x_i \in L$,
there is a natural bilinear form on $L$ given by the
integral 
\eqn\cupprod{d_{ij}=(x_i,x_j)=\int_{S} x_i \w x_j~.}  
The bilinear form on $L$ is that of the lattice $U^3 \oplus (E_8) \oplus (E_8)$.  Here,
$U$ denotes the hyperbolic lattice defined by
$\left(\matrix{0&1\cr 1 & 0}\right)$ and $A_n, D_n,
E_n$ denote the ``negative" definite lattice of the corresponding Lie
algebra root system. The form is thus given by the
negative of the Cartan matrix.

Of course, $\Nk$ and $\Nkc$ are both sublattices of $L$.  For $K3$
surfaces with a $\IZ_3$ action having only fixed points, they take
the form \refs{\arts, \taki} \eqn\Nlat{\Nk = U(3) \oplus
E_6^*(3)~,\qquad \Nkc = 
%U \oplus U(3) \oplus A_2^5  \cong 
A_2(-1) \oplus A_2^6~,} 
where for example $U(3)$ denotes the lattice with $3$ times the bilinear form of $U$
and $E_6^*$ denotes the lattice dual to the $E_6$
lattice.  Each sublattice in $\Nkc$ contains an order three discrete
symmetry action $\rho$. For the lattice $A_2$
associated with the negative Cartan matrix $\left(\matrix{-2&1\cr 1 &
-2}\right)$,  let $\{e,f\}=\{\left(\matrix{1\cr 0}\right),\left(\matrix{0\cr 1}\right)\}$ be the basis vectors which have the inner product
$(e,e)=(f,f)=-2$ and $(e,f)=1$. The $\ZZ_3$ symmetry action on the
lattice can be described by 
\eqn\kat{\rho:\quad \matrix{e &\to& f\qquad\cr f&\to& -e - f}~~~.} 
Under this action, the linear combinations $(e-\z f)$ and $(f-\z e)$ transform as
\eqn\katc{\rho:\quad \matrix{ e-\zeta f & \to & \zeta (e-\zeta
f)~\cr f-\zeta e & \to & \zeta^2 (f-\zeta e) } } 
and span an eigen-basis of two-forms with eigenvalues $\zeta$ and $\zeta^2$,
respectively.

We can now explicitly write down the complex structure $S$, or equivalently, the associated holomorphic two-form.  By \rhomalg,  $\Om^{2,0}\in \Nkc\otimes \CC\,$.  We can express $\Om^{2,0}$ as a linear combination of the basis elements of the lattice $\Nkc=A_2(-1) \oplus A_2^6$
\eqn\omexp{\Om^{2,0}= B_0 (e_0 - \z f_0)  + \sum_{i=1}^6 B_i (e_i - \z f_i) } 
where $B_0$ and $B_i$  for $i=1,\ldots,6$ are
complex constants and $\{e_0,f_0\}, \{e_i,f_i\}$ are respectively
the basis elements of $A_2(-1)$ and the six $A_2$'s.  We shall take
all pairs of $\{e,f\}$ to transform under $\rho$ as in \kat. The
constants $B_0, B_i$ determine the complex structure and
equivalently the periods with respect to a specified marking of
two-cycles on the $K3$.  They are constrained by three consistency
conditions.  The two standard ones are
\eqn\omo{\int_S \Om^{2,0} \w~ \Om^{2,0} = 0~,}
\eqn\omt{\int_S \Om^{2,0} \w ~{\bar \Om}^{2,0} =
3(|B_0|^2 - \sum_{i=1}^6 |B_i|^2)> 0~.}  
The first condition is automatically satisfied by an $\Om^{2,0}$ expressed as in \omexp. The second by itself is a weak condition and can be easily satisfied.

The third consistency condition involves the Picard lattice $N$. In
general, the Picard lattice $N$ consists of the invariant $\Nk$
and also any elements in $\Nkc$ that are $(1,1)$, {\it i.e.} $\Nkc
\cap \,\Om^\perp$.  Let $T=N^\perp$ be the complement lattice to $N$.
For a generically chosen complex structure $\Om^{2,0}$, we will have
$T=\Nkc$ and $N=\Nk$, that is all elements in $\Nkc$ will have a
$(2,0)$ and a $(0,2)$ part.  But for special complex structures, only
$T\subset \Nkc$ and $\Nk\subset N$.  Nevertheless, for elements in the
Picard lattice that transforms non-trivially under $\rho$, it
is necessary that there exists no element $h$ with $(h,h)=-2$.  For if such exists, then
by the Riemann-Roch theorem applied to $K3$ surfaces, there is an
effective divisor $\pm h$ which under the action $\rho: \pm h \to
\pm \z h {\rm ~~or~} \pm \z^2 h$ which is not possible \alexnik.
Thus, for instance, choosing $B_0=1$ and $B_i=0$ would lead to $(-2)$-curves for $h=e_i$ or  $h=f_i$ and would not be valid.  A consistent choice would be
\eqn\Bchoice{B_0=3~, B_1=1~, ~{\rm and~~~} 0<B_i\leq 1 {\rm ~~for~~} i=2,\ldots,6~.} for $B_i$ sufficiently generic.

Having established the lattice structure and the complex structure, we can now determine the possible torus curvature two-form $\om$.  As mentioned, under the discrete symmetry action, $\om$ must transform as in \ktrn.  Matching the
transformation property of $\om$ with those in \katc\ implies that
$\om$ can be written as a linear combination
\eqn\omcomb{\om=C_0 ( f_0 - \z e_0) + \sum_{i=0}^6 C_i (f_i - \z e_i)}
where $C_0$ and $C_i$'s are complex constants.  Since we require
that $\om$ does not have an $(0,2)$ component, we have the additional
constraint 
\eqn\nootwo{\int_S \Om^{2,0}\w ~\om =-3 \z ( B_0 C_0 -
\sum_{i=0}^6 B_i C_i ) = 0~.}  
Note that $\int_S {\bar \Om}^{0,2} \w
\om = 0$ is automatically satisfied for $\om$ given by \omcomb. 

For the choice of complex structure specified by \omexp\ and \Bchoice, we can for example choose $C_0=1, C_1=3$ and all other $C_i=0$ which gives
\eqn\omalgdef{\om = \om_1 + \z \om_2 = (f_0 + 3 f_1) + \z (-e _0- 3e_1) \in H^{1,1}(S)~.} 
We see that although $\om_1, \om_2 \in H^2(S,\ZZ)$ only, the combination $\om=\om_1 + \z \om_2 \in H^{1,1}(S)$ as required.   Furthermore, for $\om$ in \omalgdef\ we also have $\int_S \om \w {\bar \om} = -24$ (having set $A=\tau_2$ in \kfcon) which would give us a ${\cal N}=2$ FSY geometry with trivial gauge bundle.  

Orbifolding the FSY geometry by the $\ZZ_3$ action,  
\eqn\zthss{\rho: (\zo,\za,\zb,\zc,\zd,z)\to (\zeta^2\zo,\zeta^2\za,\zeta\zb,\zc,\zd,\zeta^2 z)~.}
the resulting supersymmetric orbifold geometry has 
$3\times 3=9$ fixed points.  (The $\ZZ_3$ action on the torus, $\rho: \th \to \zeta^2 \th$, also lead to three fixed points.)   Each fixed point is locally a $\CC^3/\ZZ_3$ orbifold that can be  minimally resolved by a $\PP^2$.  Thus the resolved manifold of this singular orbifold has Euler characteristic $\chi(\widehat Y) = (0-9)/3 + 9 \,\chi(\PP^2) = 24\,$.   For this orbifold with only $9$ fixed points, there is already a candidate local model that we can use to resolve each singularity of the geometry.  This is the well-known local Calabi-Yau metric that resolves $\CC^3/\ZZ_3$ \refs{\Calabi, \StrWi}.  However, careful analysis is required to verify that such gluing-in can preserve the required supersymmetry and anomaly conditions.

\newsec{Discussions}

We have constructed new non-K\"ahler heterotic geometries $Y=
X/\Gamma$ by orbifolding the torus bundle over $K3$ base geometry by
a discrete symmetry group $\Gamma$.
 The orbifolds $Y$ are clearly topologically distinct from the starting torus bundle geometry $X$.
  For smooth orbifold models, the first fundamental group $\pi_1(Y)$ additionally contains the
symmetry group $\Gamma$.  For singular orbifolds where $G$ has a fixed locus set, the
resolution of singularities will generically give a non-zero Euler characteristic for the
 resolved manifold $\widehat{Y}$.

The orbifold models we have constructed can be supersymmetric or
break all supersymmetries.  We have given explicit constructions of
orbifold solutions where the $K3$ base of the FSY model is either a
Kummer $K3$ or has a branched covering description or a more general
algebraic description. Those orbifold geometries arising from a
shift action on the fiber can still be viewed as having a fiber
space structure. The base is now the $K3$ orbifolded by the
 discrete action and the complex structure of the torus jumps along the fixed locus
 set of the discrete action.  Rotation actions on the torus
  fiber may result in orbifold models with singularities, which
  potentially can have a non-vanishing Euler characteristic as well as a
  non-vanishing number of generations.
  The smooth resolution of these non-K\"ahler singular orbifold solutions is an
important question that shall be addressed in future work.

The smooth non-supersymmetric geometries we have constructed are each an elliptic fibration over a complex surface.  For example, the one in Section 5.1 utilizing a branched covered $K3$ surface is an elliptic fibration over a $\PP^1 \times \PP^1$ base.  Being non-supersymmetric, the holomorphic three-form of the FSY geometry has been projected out.  It is thus interesting to ask whether these smooth torsional elliptically fibered three-folds more generally can support a no-where vanishing holomorphic three-form.  Such would lead directly to smooth supersymmetric heterotic geometries.   Some preliminary metric ans\"atze for these elliptically fibered three-folds has been proposed in \unpub.  Interestingly, from the perspective of F-theory and heterotic string duality \refs{\MoVa,\becktwo}, elliptically fibered three-folds should be dual to F- or M-theory flux compactifications on Calabi-Yau four-folds \KM\ that are $K3$ fibrations.   Recent work using this approach appears to indicate the existence of such class of heterotic flux compactifications \bese.

It is worthwhile to emphasize that the FSY geometries, at least for
those that preserve ${\cal N}=2$ SUSY, have a conformal field theory
description \adel\ and represent a class of string vacua valid to
all orders in $\alpha'$.  The construction of geometric orbifold
models described here should therefore have an analogous orbifold
description from the CFT perspective.  The CFT orbifold models that
are constructed from freely acting discrete groups should just
consist of a projection. For the non-freely acting ones, the
resolution of singularities should correspond to the addition of
twisted sector modes.  It would be interesting to work out the CFT
description of the orbifold models constructed herein.

An important question which we leave for future work is whether the
non-super-symmetric orbifolds are stable in the $g_s$ loop expansion.
Some beautiful examples of non-supersymmetric stable orbifolds were
constructed in \kasi. Non-supersymmetric unstable orbifolds can lead
to an interesting decay process as has been discussed in the recent
literature \refs{\tachyonI,\tachyonII}. In particular in \tachyonII\
non-compact non-supersymmetric orbifolds were shown to decay into
supersymmetric ALE spaces. It would be interesting to see if an
analysis along the lines of \tachyonI\ can be performed for the
heterotic orbifold models we have constructed.

\bigskip\bigskip\bigskip
\centerline{\bf Acknowledgements}
\medskip
We thank A.~Adams, A.~Bergman, K.~Becker, R.~Donagi, J.-X.~Fu,
S.~Kachru, B.~Lian, J.~Lapan, D.~Morrison, K.~Oguiso, A.~Sarti, A.~Strominger,
A.~Subotic, A.~Todorov, and C.~Vafa for useful discussions and correspondence.
M.~Becker would like to thank the Harvard Department of Physics for
hospitality where part of this work was carried out.   M.~Becker and L.-S.~Tseng would also like to thank the hospitality of the 6th-Simons Workshop in Mathematics and Physics at SUNY Stony Brook near the completion of this work.  M.~Becker is supported by NSF grant PHY-0505757 and the University of Texas A\&M.  L.-S.~Tseng is supported in part by NSF grant PHY-0714648.  S.-T.~Yau is supported in part by NSF grants DMS-0306600 and PHY-0714648.

\bigskip\bigskip\bigskip

\listrefs

\end